\documentstyle[aps,preprint]{revtex}
\begin{document}
\draft 
\preprint{ISSP \today}
\title
{
Quasi-particle spectra 
around a single vortex
in a d-wave superconductor
}
\author{Y. Morita$^{1}$, M. Kohmoto$^{1}$ and K. Maki$^{2}$}
\date {\today}
\address
{
$^{1}$Institute for Solid State Physics,
University of Tokyo
7-22-1 Roppongi Minato-ku, Tokyo 106, Japan
}
\address
{ 
$^{2}$Department of Physics and Astronomy,
University of Southern California Los Angeles,
Cal. 90089-0484, USA
}
\maketitle
\begin{abstract} 
Using
the Bogoliubov-de Gennes equation, 
we study 
the quasi-particle spectra 
around a single vortex
in a d-wave superconductor,
where
a magnetic field is parallel to the c-axis.
In the temperature region 
where the Ginzburg-Landau theory is valid,
we find that 
the local density of states preserves a circular symmetry
when the symmetry of 
the superconducting order parameter is  {\it pure} d-wave.
It, however, exhibits a four-fold symmetry
when 
the mixing of a s-wave component occurs.
A peak 
with a {\it large energy gap}
is found
in the local density of states
at the center of the vortex,
which corresponds to the lowest bound state.
Our results are consistent 
with a recent scanning tunneling microscopy experiment 
in an YBa$_{2}$Cu$_{3}$O$_{7-{\delta}}$ (YBCO) monocrystal.
The breakdown of 
the Eilenberger theory
in YBCO in particular and 
in the high-T$_{c}$ superconductors in general
is discussed.
\end{abstract}  
\pacs{}
\narrowtext
It is now well established that
most of the properties
of the high-$\rm T_{c}$ superconductors
are described in terms of a d-wave superconductor
\cite{exp1,exp2}
with possible exception of 
the electron-doped 
Nd$_{2-x}$Ce$_{x}$CuO$_{4}$
\cite{exp3}. Therefore 
it is of great interest
to study  vortex states
in a d-wave superconductor
\cite{maki1,maki2,maki3}.
For example,
in a magnetic field parallel to the c-axis,
it was 
predicted that a square lattice of vortices
tilted by ${\pi}/4$ from the a-axis
is the most stable for $T<0.8T_{c}$\cite{maki2},
which has been seen recently by 
a small angle neutron scattering \cite{neutron} 
and a scanning tunneling microscopy (STM) experiment \cite{geneve}
in YBa$_{2}$Cu$_{3}$O$_{7-{\delta}}$ (YBCO) monocrystals
at low temperatures ($T<10$K),
although elongated in the b direction. 
One of the interpretations 
is that this distortion is due to 
the orthorhomibicity of YBCO,
although there are alternative interpretations
based on the (d+s) admixture \cite{sd1,sd2}.

Here we shall restrict ourselves 
to a single vortex
in a d-wave superconductor, 
where a magnetic field is parallel to the c-axis.
Early analyses 
of 
the quasi-particle spectra
around a single vortex
in a d-wave superconductor
within 
the Eilenberger theory
(a semi-classical theory of a superconductor)
\cite{eilen} 
exhibits a surprising fourfold symmetry
in the local density of states
\cite{maki4,kyoto},
which has not been seen experimentally\cite{geneve}.
These analyses rely on the Riccati equation 
which  is obtained in \cite{maki4} by simplifying
the Eilenberger equation\cite{eilen}.
It is to be noted that, in \cite{kyoto},
the low-temperature region,
where the Ginzburg-Landau (GL) theory is invalid,
is also investigated 
by solving the Eilenberger equation self-consistently. 

We believe that the above Eilenberger theory, 
which is essentially a semi-classical approximation,
is invalid for the high-T$_{c}$ superconductors\cite{maki5}.
In order to clarify the point,
we shall give an analysis of the experimental results.

In YBCO \cite{geneve},
no peaks 
are found
in 
zero-bias
tunneling conductance  
at the center of the vortex,
in contrast to
a s-wave superconductor NbSe$_{2}$ \cite{sstm}.
Further, at the center of the vortex,
there is a peak 
around $E=0.25 \Delta $
in the tunneling conductance,
where $\Delta$=280K is 
the superconducting order parameter 
of YBCO at $T=0$K.
Then
the most natural interpretation is
that
this corresponds to the lowest bound state
for  a vortex 
in a d-wave superconductor
analogous to the one predicted by
Caroli, de Gennes and Matricon \cite{degennes}.
Since the lowest bound-state energy is given
approximately by $E_0 = {\Delta}/{\pi}({p_{F}}\xi)^{-1}$
\cite{maki4,sbdg1},
where $p_{F}$ and $\xi$ are 
the Fermi momentum and the coherence length
respectively,
this implies ${p_{F}}{\xi}\sim 1$ in YBCO.
This is also consistent with the chemical potential of YBCO deduced
from the analysis of the spin gap seen in an inelastic neutron
scattering experiment from monocrystals of YBCO \cite{RM,maki6}. 

Therefore in order to describe the quasi-particle spectra 
around a single vortex in high-T$_c$ superconductors, 
it is crucial to study the Bogoliubov-de Gennes equation
which is given by
\begin{eqnarray} 
&&
\Bigl \{  
-{\frac{1}{2m}}
({\nabla}-ie{\bf {A}}({\bf x}))^{2}-\mu
\Bigr \}
u_{n}({\bf x};\hat{k})
+
{\Delta}({\bf x};\hat{k})
v_{n}({\bf x};\hat{k})
\nonumber
\\
&&=
{\epsilon }_{n}
(\hat{k})
u_{n}({\bf x};\hat{k}), 
\nonumber
\\
&&
-
\Bigl \{  
-{\frac{1}{2m}}
({\nabla}+ie{\bf {A}}({\bf x}))^{2}-\mu
\Bigr \}
v_{n}({\bf x};\hat{k})
+
{{\Delta}^{*}({\bf x};\hat{k})}
u_{n}({\bf x};\hat{k})
\nonumber
\\
&&=
{\epsilon }_{n}
(\hat{k})
v_{n}({\bf x};\hat{k}),
\label{bdg1} 
\end{eqnarray}
where $u_{n}({\bf x};\hat{k})$ 
and $v_{n}({\bf x};\hat{k})$
are
the quasiparticle amplitudes,
${\Delta}({\bf x};\hat{k})$
is the pair potential,
${\bf {A}}({\bf x})$ is the vector potential
which is neglected assuming $H{\ll}H_{c2}$
\cite{degennes},
and
$\mu$ is the chemical potential
which is identified with the Fermi energy.
In this paper,
we consider
two types of the pair potential
${\Delta}({\bf x};\hat{k})$,
which are 
'{\it pure} d-wave' {\cite{maki4}}{\cite{ren1}}  
\begin{equation}
{\Delta}({\bf x};\hat{k})
={\Delta}\tanh (r/{\xi }){\cos}(2\theta(\hat{k}))e^{i{\phi}},
\label{test1}
\end{equation}
where
$r$, $\phi$ and $\theta(\hat{k})$
are defined by
${\bf x}=(r\cos(\phi),r\sin(\phi))$,
$({\hat{k}}_{x},{\hat{k}}_{y})
=({\cos}(\theta(\hat{k})),{\sin}(\theta(\hat{k})))$,
$\Delta$ is a real constant
and
$\xi $ is the coherence length of the superconductor;
and '(d+s)-wave' {\cite{ren1}} 

\begin{eqnarray}
{\Delta}({\bf x};\hat{k})
&=&
{\Delta}\tanh(r/{\xi})\cos(2\theta(\hat{k}))e^{i{\phi}}
+
b_{0}(r/{\xi})e^{-i{\phi}}
\nonumber
\\
&&{\rm for}\ 0<r {\le}{\xi},
\nonumber
\\
{\Delta}({\bf x};\hat{k})
&=&
({\Delta}\tanh(r/{\xi})+a_{0}({\xi}/r)^{2})
\cos(2\theta(\hat{k}))e^{i{\phi}}
\nonumber
\\
&&+
({\xi}/r)^{2}{\cdot }
\nonumber
\\
&&\cos(2\theta(\hat{k}))
{\bigl (}
a_{-1}e^{-3i{\phi}}+a_{1}e^{5i{\phi}}
{\bigr )}
\nonumber
\\
&&+
({\xi}/r)^{2}{\cdot }
\nonumber
\\
&&
{\bigl (}
b_{0}e^{-i{\phi}}+b_{1}e^{3i{\phi}}
{\bigr )}
\nonumber
\\
&&{\rm for}\ {\xi}<r.
\label{test2}
\end{eqnarray}

The pair potentials 
are obtained 
by the GL theory
for an anisotropic superconductor\cite{maki4,ren1}
and applicable at not too low temperatures
(an estimate of the temperature region 
where the GL theory is valid is
$[0.5T_{c},T_{c}]$ for the 
s-wave superconductors,
when one consider the quasi-particle spectra 
around a single vortex \cite{sbdg1}).

In writing (\ref {bdg1}), 
we neglected 
the 
{\it non-commutability
between $\hat k$ and $\bf x$}
{\cite {bruder}}{\cite{sch}}. 
The correction 
is $O(1/p_{F}{\xi})$
and 
irrelevant
at least in the study of 
systems with a long coherence length
e.g.
a superconducting phase
of a heavy-fermion system 
($p_{F}{\xi}\sim 10$)
and
${}^{3}$He superfluidity
($p_{F}{\xi}\sim 100$),
but may have a serious influence 
in the study of 
the high-$\rm T_{c}$ superconductors,
where $p_{F}{\xi}\sim 1$ as is discussed above.
But our results 
seem to be consistent 
with experimental results of YBCO 
as will be discussed below. 
Here it is to be noted that, 
in contrast to 
a usual semi-classical theory of a superconductor
(the Eilenberger theory),
our method takes into account
quantization of the angular momentum.
In fact, as we shall see,
the quasi-particle spectra 
is totally different from ones
obtained semi-classically\cite{maki4,kyoto}. 

In order to solve the Bogoliubov-de Gennes equation (\ref{bdg1})
numerically,
it is convenient to expand
the quasi-particle amplitudes
$u_{n}({\bf x};\hat{k})$
and
$v_{n}({\bf x};\hat{k})$
as 
\begin{eqnarray}
&&
u_{n}({\bf x};\hat{k})
=
{\sum_{l=-\infty}^{\infty}}
{\sum_{j=1}^{\infty}}
u_{n,l,j}(\hat{k})\ {\psi}_{j,|l|}(r)\exp(il{\phi}),
\nonumber
\\
&&
v_{n}({\bf x};\hat{k})
=
{\sum_{l=-\infty}^{\infty}}
{\sum_{j=1}^{\infty}}
v_{n,l,j}(\hat{k})\ {\psi}_{j,|l-1|}(r)\exp(i(l-1){\phi}).
\label{mode}
\end{eqnarray}
Here
${\psi}_{i,\nu}(x)=
\frac{1}{{\sqrt{2\pi}}RJ_{\nu+1}(\alpha _{i,\nu})}
J_{\nu}(\alpha _{i,\nu}x/R)$
($J_{\nu}(x)$
is the Bessel function),
$\alpha _{j,\nu}$ is the $j$ th positive zero point
of $J_{\nu}(x)$
and $R$ is the radius of the system.
In our numerical calculation,
the number of the basis
is about 80000 for {\it pure} d-wave 
and about 4800 for (d+s)-wave.
It is sufficient to discuss physical properties
around a single vortex in a d-wave superconductor.

The quantity of interest
for comparison to STM experiments
is a local density of states in a superconductor
\begin{eqnarray}
N(E,{\bf x})
&&=
\sum_{{\hat k},n}
[\ 
|u_{n}({\bf x},{\hat k})|^{2} 
\ \delta (E-{\epsilon _{n}({\hat k})})
\nonumber
\\
&&+
|v_{n}({\bf x},{\hat k})|^{2}
\ \delta (E+{\epsilon _{n}({\hat k})})
].
\label
{ldos}
\end{eqnarray}

The local densities of states 
$N(E,{\bf x})$ 
for {\it pure} d-wave 
are shown in
Figs. 1-4
for $r/{\xi}=$0.0, 1.0, 2.0 and 10.0,
respectively.
The parameters are chosen as
$p_{F}\xi =1.41$
and
$2m{\xi}^{2}{\Delta}=2.82$,
from the experimental condition\cite{geneve}.
We also set $R/{\xi }=40$.
One can see 
peaks 
in the local densities of states
corresponding to 
bound states
around a single vortex,
which are localized near the vortex center. 

For $r/\xi=0.0\ {\rm and}\ 1.0$ (Figs. 1 and 2),
a peak with a {\it large energy gap},
which corresponds to the lowest bound state,
appears in the local density of states.
The peak has a width
due to the internal degree of freedom
in the $\hat k$ space.
The result is consistent with a recent STM experiment\cite{geneve}.
For $r/\xi=10.0$ (Fig. 4),
$N(E,{\bf x})$'s around a vortex and without a vortex
are almost identical.
This indicates that 
the bound state is localized within this distance.
We believe 
that
the large energy gap is closely related to
the quantization of the angular momentum.
We note that
there is 
a clear asymmetry 
in the local density of states
at $E$ and $-E$
due to the formation of the bound states
\cite{Mac}.

The {\it integrated}
local densities of states 
near the zero energy
$\int _{0}^{0.17\Delta }dE\ N(E,{\bf x})$
are shown in Figs.5 and 6 for
{\it pure} d-wave and (d+s)-wave respectively.
The parameters are chosen as
$p_{F}\xi =1.33$,
$R/{\xi }=30$,
$2m{\xi}^{2}{\Delta}=2.82$,
$2m{\xi}^{2}a_{m}=0.05\ (m=-1,0,1)$
and
$2m{\xi}^{2}b_{m}=0.05\ (m=0,1)$.
The choice of high energy cutoff
does not change the results quantitatively.
Note that
the local density of states preserves a circular symmetry
for {\it pure} d-wave.
On the other hand,
it exhibits a clear four-fold symmetry
when 
the mixing of a s-wave component occurs
( see Fig. 6 ).

In conclusion,
we have investigated
the quasi-particle spectra
around a single vortex in a d-wave superconductor
by solving the Bogoliubov-de Gennes equation
numerically.
The local densities of states
are consistent with a recent STM experiment\cite{geneve}.
The result shows
a peak with a {\it large energy gap}
in the local density of states
at the center of the vortex,
which corresponds to the lowest bound state.
We find that 
the local density of states preserves a circular symmetry
when the symmetry of 
the superconducting order parameter is  {\it pure} d-wave.
It, however, exhibits a four-fold symmetry
when 
the mixing of a s-wave component occurs.
However 
it is possible that
a four-fold symmetry
of a local density of states
without a mixing of a s-wave component
\cite{maki4,kyoto}
appears
due to 
the
{\it noncommutability
between $\hat k$ and $\bf x$}.
A fully quantum-mechanical
treatment
of the Bogoliubov-de Gennes equation
is left as a future problem.
Also
a self-consistent treatment 
is needed
in low temperatures,
where the GL theory is invalid.
Moreover,
a treatment of 
a vortex $\it lattice$
is essential 
when one investigates
the thermodynamic quantities.

One of the authors (K.M.)
thank 
$\emptyset$. Fischer
and
Ch. Renner
for illuminating discussions on their STM results.
This work is in part supported by
National Science Foundation
under grant number
DMR92-18371 and DMR95-31720
and by
Grant-in-Aid for Scientific Research
from the Ministry of Education,
Science and Culture, Japan.

\begin{figure}

Fig. 1 
The local density of states 
$N(E,{\bf x})$
around a vortex ($\diamond $)
and that without a vortex ($+$),
where $r/{\xi}=0.0$.
The rescaled energy 
is defined by 
$E/{\Delta}$.

Fig. 2
The local density of states 
$N(E,{\bf x})$
around a vortex ($\diamond $)
and that without a vortex ($+$),
where $r/{\xi}=1.0$.
The rescaled energy 
is defined by 
$E/{\Delta}$.

Fig. 3
The local density of states 
$N(E,{\bf x})$
around a vortex ($\diamond $)
and that without a vortex ($+$),
where $r/{\xi}=2.0$.
The rescaled energy 
is defined by 
$E/{\Delta}$.

Fig. 4
The local density of states 
$N(E,{\bf x})$
around a vortex ($\diamond $)
and that without a vortex ($+$),
where $r/{\xi}=10.0$.
The rescaled energy 
is defined by 
$E/{\Delta}$.
Weak anisotropy 
appears due to boundary effect. 

Fig. 5
The {\it integrated} local density of states 
$\int _{0}^{0.17\Delta }dE\ N(E,{\bf x})$
for {\it pure} d-wave (left)
and that for (d+s)-wave (right),
where $\phi =3\pi /8$.
The lines are guide for eyes.

Fig. 6
The {\it integrated} local density of states 
$\int _{0}^{0.17\Delta }dE\ N(E,{\bf x})$
for {\it pure} d-wave ($\diamond $)
and that for (d+s)-wave ($+$),
where $r/{\xi}=3.0$.
The lines are guide for eyes.

\end{figure}

\end{document}